\documentclass[12pt,a4paper]{article}
\usepackage{amsfonts}
\newcommand{\RR}{\mathbb{R}}
\newcommand{\CC}{\mathbb{C}}
\newcommand{\ZZ}{\mathbb{Z}}

\newcommand{\io}{\iota}
\newcommand{\bio}{\bar{\io}}
\newcommand{\ep}{\epsilon}
\newcommand{\frc}[2]{{\textstyle\frac{#1}{#2}}}
\newcommand{\one}{1\kern-3pt{\rm l}}
\newcommand{\hf}{\frc{1}{2}}
\newcommand{\bw}{{\textstyle\bigwedge}}
\newcommand{\Wp}{{\bw^+}}
\newcommand{\Wm}{{\bw^-}}
\newcommand{\br}[1]{\langle #1\rangle}
\newcommand{\arr}[1]{\,\smash{\mathop{\longrightarrow}\limits^{#1}}\,}
\newcommand{\sco}[1]{\,[\![\,{#1}\,]\!]\,}
\newcommand{\nein}{\in\kern-0.8em/\,}
\newcommand{\id}{\mbox{id}}
\newcommand{\Ad}{\mbox{Ad}\,}

\newcommand{\En}{\mbox{End}}
\newcommand{\End}{\En\,}
\newcommand{\Hom}{\mbox{Hom}\,}
\newcommand{\ad}{\mbox{ad}\,}
\newcommand{\Star}{$\phantom{|}^*$}
\newcommand{\sad}{\mbox{\scriptsize{ad}}}
\newcommand{\sAd}{\mbox{\scriptsize{Ad}}}
\newcommand{\skt}{{\hat\otimes}}
\newcommand{\Spin}{\mbox{Spin}\,}
\newcommand{\Spinc}{\mbox{Spin}_\CC\,}
\newcommand{\llra}{\relbar\joinrel\relbar\joinrel\longrightarrow}
\newcommand{\mapr}[1]{\smash{\mathop{\llra}\limits^{#1}}}
\newcommand{\mapd}[1]{\Big\downarrow\rlap{$\vcenter{\hbox{$\scriptstyle#1$}}$}}

\begin{document}
%\vspace*{-2cm}
%\rightline{math-ph/9908029}
%\vspace{2cm}
\begin{center}
 {\LARGE\bf An Introduction\\[3mm] to Clifford Supermodules}\\[15mm]
  {\large G.\ Roepstorff and Ch.\ Vehns}\\
  Institute for Theoretical Physics\\
  RWTH Aachen\\
  D-52062 Aachen, Germany\\
  e-mail:\\
  roep@physik.rwth-aachen.de\\
  vehns@physik.rwth-aachen.de\\[2cm]
  {\bf Abstract}
\end{center}
\begin{quote}
We analyze the Clifford action on superspaces with a view on generalized
Dirac fields taking values in some Clifford supermodule. The stress here
is on two principles: complexification and polarisation. For applications
in field theory, the underlying vector space may carry either a Euclidean
or a Minkowskian structure.
\end{quote}

\section{Introduction}

This is the first part of a series of articles in which we aim to present
those ideas and methods which can now be effectively used by physicists
working in the theory of fermions. The reason for studying the subject
is the observation that generalized (multi-component) Dirac fields are
instances of Clifford supermodules [1,2]. The particular point of view 
presented here will help to better understand the role of the Higgs field 
and Yukawa interactions in gauge theories, seemingly the essential
features of the Standard Model. In pursuing this idea, we analyze 
the mathematical structure underlying the modern theory of Dirac operators 
on manifolds [3] and their relation to superconnections on $\ZZ_2$-graded 
vector bundles with an outlook on possible applications in particle physics. 
We introduce only tools appropriate for a consistent and clear formulation
of field theoretical models
and begin by discussing the algebraic fundament which is the structure of 
Clifford algebras [4,5,6] and their action on superspaces. 
The emphasis will be on two important principles, {\em complexification and
polarisation}, and on two structural aspects, {\em skew tensor products and
twisting}, thereby stressing the superalgebra point of view.
But to keep both feet on the ground we
complement the discussion by studying specific examples for illustration.
The ideas, though presented here in an abstract setting, will
come alive when applied to supersymmetric Dirac theory [7] and the gauge 
theory of quarks and leptons [2].

While spinors belong to the standard repertoire of physicists since
Dirac's seminal work on spin $\hf$ particles, 
in the mathematics community, the use of spinors has been spreading slowly.
It is only recent that spinor modules and the associated Dirac operators
have developed into a fundamental tool in differential geometry of
Riemannian (spin) manifolds. With same emphasis, Penrose [8], Witten [9], and
Connes [10] place spinors at the fundament of their constructions in
geometry, be it commutative or non-commutative.

\section{Clifford Algebras}

There are various ways to naturally associate an algebra to some 
$n$-dimen\-sional real vector space $V$. Each one is characterized 
by a universal property and obtained explicitly from the tensor
algebra $T(V)$ by taking the quotient with respect to some 
two-sided ideal $I$. The Clifford algebra provides a prominent example.
In more detail:
let $q(v)$ denote a (possibly degenerate) quadratic form in $V$, or,
equivalently, let there be given a bilinear symmetric form $(,)$ in $V$
such that $q(v)=(v,v)$. 
Let $I$ be the ideal in $T(V)$ generated by the elements $v\otimes v+q(v)1$
($v\in V$) where 1 is the unit in $T(V)$. Then the Clifford algebra $C(V)$ 
over $V$ is the quotient $T(V)/I$. The embedding $V\to C(V)$ allows us to
identify $v\in V$ with $v\in C(V)$ without confusion.
By construction, the {\em Clifford relation\/}\\[2mm]
\begin{equation}
  \label{eq:ideal}
                  v\otimes v+q(v)1=0\qquad (v\in V)
\end{equation}
holds in in any Clifford algebra $C(V)$. In addition, it has the following 
universal property.
Any linear map $c: V\to A$ into an associative algebra $A$ with unit
satisfying 
\begin{equation}
  \label{eq:clm}
                   c(v)^2+q(v)1=0
\end{equation}
extends to an algebraic homomorphism $c:C(V)\to A$. A linear map $c$
satisfying (\ref{eq:clm}) is called a {\em Clifford map}.

The first thing to be noticed about this construction is that the 
$\ZZ$-grading of the tensor algebra $T(V)$ is lost when the quotient $T(V)/I$ 
is formed. Nevertheless, there remains a $\ZZ_2$-graded structure given by the 
involution $\tau(v)= -v$ in $V$ (leaving the ideal $I$ invariant). 
With respect to the eigenvalues $\pm1$ of $\tau$ we may write
$$
               C(V) = C^+(V)\oplus C^-(V).
$$
One remark is in order. We shall often use the parities $\pm1$ instead of
the numbers $0,1\in\ZZ_2$ to indicate the grading, because it seems more
convenient. Also, parities are more intuitive from the physics point of view.
The grading gives $C(V)$ the structure of a {\em superalgebra} as the
following properties demonstrate:
$$
     C^\pm(V) C^+(V)\subset C^\pm(V),\qquad
     C^\pm(V) C^-(V)\subset C^\mp(V).
$$
As with any superalgebra, the positive part $C^+(V)$ is a subalgebra.

Second, from (\ref{eq:ideal}) we immediately obtain the 
anticommutation relation
\begin{equation}
  \label{eq:vv}
             vv'+v'v+2(v,v')1 =0\qquad(v,v'\in V)
\end{equation}
valid in the Clifford algebra $C(V)$. For a basis $\{e_i\}_{i=1}^n$ in $V$
we tacitly assume that $(e_i,e_k)=0$ if $i\ne k$. Of course, this condition
would be empty on a subspace where the bilinear form vanishes. Now, if $I$ 
runs over all subsets of $\{1,\ldots,n\}$, then (\ref{eq:vv}) can be used 
to show that the ordered products 
\begin{equation}
  \label{eq:bas}
   e_I= e_{i_1}\cdots e_{i_k}\qquad(i_1<\cdots<i_k),
\end{equation}
where $I=\{i_1,\ldots,i_k\}$, form a basis in $C(V)$. For the empty set, one
puts $e_\emptyset=1$. This then demonstrates that
\begin{eqnarray*}
   {\rm dim\,}C^+(V) &=& \sum_{k={\rm even}}{n\choose k}=2^{n-1}\\
   {\rm dim\,}C^-(V) &=& \sum_{k={\rm odd}}{n\choose k}=2^{n-1}.     
\end{eqnarray*}
Note that the dimension $2^{n-1}$ of $C^\pm(V)$ is independent of
the choice of the bilinear form $(,)$, be it degenerate or not.

Third, if the vector space $V$ has not been equipped with a bilinear form,
we express this by saying that $(,)$ vanishes everywhere in $V$. 
The Clifford algebra $C(V)$ then coincides with the exterior algebra $\bw V$.
In this limiting case, the exterior algebra inherits the $\ZZ$-graded structure
of the tensor algebra $T(V)$. Elements $a\in\bw^kV$ are said to be
homogeneous of degree $k$ or are simply called $k$-vectors.
We shall however be mainly concerned with the $\ZZ_2$-grading of $\bw V$: 
$$
   \Wp V =\sum_{k={\rm even}}\bw^kV,\qquad \Wm V =\sum_{k={\rm odd}}\bw^kV\ .
$$
It is common practice to prefer the notation $a\wedge b$ over $ab$ for the
product of two elements $a,b\in\bw V$. We shall not always adhere to this
convention and warn the reader.
As we shall see lateron, there is a close relationship between
Clifford algebras and their associated exterior algebras. Thus, to avoid 
confusion it seems wise to follow the tradition and to distinguish between the
Clifford product $ab$ and the exterior product $a\wedge b$.

\section{Clifford Modules and Supermodules}

One example, intensively studied and applied by physicists, is provided by the
four-dimensional Minkowski space $M_4$ with the Lorentz metric.
Dirac's idea of representing the Clifford algebra $C(M_4)$ by complex 
$4\times 4$ matrices can be generalized and leads us to the concept of
a {\em Clifford module}. To start with, we introduce the concept of
a {\em real\/} Clifford module, which is a real vector space $E$ together
with an algebraic homomorphism
\begin{equation}
  \label{eq:cm}
                c:C(V)\to{\rm End\,}E.
\end{equation}
By the universal property of $C(V)$, it suffices to assume that there be
given a linear map $c:V\to{\rm End\,}E$ satisfying the Clifford relation
$c(v)^2+q(v)1=0$.

Of particular interest is the case where $E$ is $\ZZ_2$-graded (hence is a
superspace) giving End$\,E$ the structure of a superalgebra:
\begin{eqnarray*}
  {\rm End^\pm}E &=& {\rm Hom}(E^+,E^\pm)\oplus{\rm Hom}(E^-,E^\mp)\\
  {\rm End}\,E   &=& {\rm End^+}E\oplus{\rm End^-}E.
\end{eqnarray*}
We would then require that the map (\ref{eq:cm}) is in fact
a homomorphism between superalgebras (respecting the grading):
$$
               c:C^\pm(V)\to{\rm End^\pm}E.
$$
We say that $E$ is a {\em Clifford supermodule}.
For this to be the case it suffices to assume that $c(v)\in{\rm End^-}E$
(all $v\in V$), i.e., that $c(v)$ changes the parity of vectors in $E$:
$$
                c(v): E^\pm \to E^\mp.
$$
There are further conditions that may be imposed on Clifford modules.
We mention two of them.\\
(1) If the Clifford algebra acts irreducibly on $E$, i.e., if $c:C(V)\to 
{\rm End}\,E$ is an isomorphism, $E$ is called an {\em irreducible
Clifford module}. The module $E$ is reducible if it contains an invariant
(proper) subspace, completely reducible if it is a direct sum of invariant
subspaces.\\
(2) There is a linear operation on $C(V)$ 
that respects the grading and mimics the notion of {\em passing to the 
adjoint}, familiar from operator algebras. In fact, we can extend the 
involution $v^*=-v$ in $V$ to all of $C(V)$ so as to satisfy the axioms
of a \Star algebra:
\begin{eqnarray*}
      1^*    &=& 1\\
      a^{**} &=& a\\
      (ab)^* &=& b^*a^*,\qquad a,b\in C(V).
\end{eqnarray*}
Frequently, ${\rm End\,}E$ is a \Star algebra, in which case it seems natural
to require that $c(a)^*=c(a^*)$.
The Clifford module $E$ is then said to be {\em selfadjoint}.
This entails that there is some non-degenerate symmetric bilinear
form $(,)$ in the Clifford module $E$ and, for any $A\in{\rm End}E$, one 
defines the adjoint $A^*$ with respect to $(,)$. Supposing moreover that $E$
is a Clifford supermodul, we would have to add the condition 
$({\rm End}^\pm E)^*={\rm End^\pm}E$. The following remark is useful in
applications. Given a Clifford map $c:V\to {\rm End}\,E$. In order to
establish selfadjointness of the extension $c:C(V)\to{\rm End}\,E$ it
suffices to check that the Clifford map satisfies $c(v)^*+c(v)=0$.

\section{The Exterior Algebra $\bw V$ as Clifford\\ Supermodul}

As before, $V$ is assumed to be an $n$-dimensional real vector space.
With $V^*$ its dual,
there is a canonical quadratic form $q$ in $V^*\oplus V$ given by
$$    
       q(u,v)=u(v),\qquad u\in V^*,v\in V
$$
and a Clifford algebra $C(V^*\oplus V)$ with the canonical embedding
\begin{equation}
  \label{eq:emb}
  V^*\oplus V \to C(V^*\oplus V),\quad (u,v)\mapsto \ep(v)-\io(u).
\end{equation}
We have thus introduced $\ep(v)$ and $\io(u)$ as (odd) generators
of the Clifford algebra. They satisfy the relations
\begin{equation}
  \label{eq:car}
\{\io(u),\ep(v)\}=u(v)1,\qquad\{\io(u),\io(u')\}=0,\qquad
\{\ep(v),\ep(v')\}=0 
\end{equation}
known in physics as {\em canonical anticommutation relations\/} (CAR).

Conversely, suppose $V$ has dimension $2n$ and subspaces $V_\pm$ of
dimension $n$, dual to each other in the sense that $V_\pm\cong V_\mp^*$, 
such that
\begin{equation}
  \label{pol}
           V\cong V_+^*\oplus V_+ \ .    
\end{equation}
We say that the isomorphism (\ref{pol}) provides a (real) {\em polarisation\/}
of $V$ if the quadratic form $q$ in V corresponds to the canonical quadratic 
form in $V_+^*\oplus V_+$ up to a factor $\hf$, i.e.,
$$
     (v_-,v_+)=\hf q(v),   \qquad v=v_++v_-  \qquad ( v_\pm\in V_\pm).
$$
The factor $\hf$ is automatic. It follows from $(v_\pm,v_\pm)=0$ and 
$$
      q(v)=(v_++v_-,v_++v_-)=(v_+,v_-)+(v_-,v_+)=2(v_-,v_+).
$$

{\bf Example}. The two-dimensional Minkowski space, $M_2$, admits a
polarisation because the Lorentz metric may be written
$$
             (x,x)=(x^0-x^1)(x^0+x^1).
$$
The subspaces are $V_\pm=\{x\in M_2\ |\ x^0\pm x^1=0\}\cong\RR$.

The intimate connection between the concepts of a Clifford and
an exterior algebra becomes apparent when we now show that there
is a natural isomorphism between superalgebras,
\begin{equation}
  \label{eq:iso}
 C(V^*\oplus V)\cong\End\bw V,
\end{equation}
or phrased differently, that the above CAR algebra is irreducibly represented
on the superspace $\bw V$. 

The proof proceeds in steps. First, the linear map $\ep:V\to C(V^*\oplus V)$
satisfying $\ep(v)^2=0$ extends to an embedding of algebras,
$$
               \ep:\bw V\to C(V^*\oplus V),
$$
which allows us to identify $\bw V$ with $\ep(\bw V)1$. This assigns to $\ep(v)$
the role of a {\em multiplication operator\/} on $\bw V$,
$$
             \ep(v)a = v\wedge a\qquad(v\in V,\ a\in\bw V),
$$
and to $\io(u)$ the role of a {\em contraction operator}, uniquely
characterized by the conditions
\begin{eqnarray*}
 \io(u)1&=&0\\
 \io(u)(v\wedge a)&=&u(v)a-v\wedge\io(u)a\qquad(u\in V^*,\ v\in V,\ a\in\bw V).
\end{eqnarray*}
Second, the algebra $\End\bw V$ is easily shown to be generated by the 
operators $\ep(v),\io(u)$ and hence by the operators $\ep(v)-\io(u)$.
But (\ref{eq:emb}) says that the elements $\ep(v)-\io(u)$ also generate
the Clifford algebra $C(V^*\oplus V)$ and so the Clifford map
$$
    V^*\oplus V\to\End\bw V,\qquad (u,v)\mapsto\ep(v)-\io(u),
$$
extends to an isomorphism (\ref{eq:iso}). Since both $\ep(v)$ and $\io(u)$
are operators on $\bw V$ of odd parity, it is guaranteed that the isomorphism
respects the $\ZZ_2$-grading. To summarize, the exterior algebra $\bw V$ is
an irreducible Clifford supermodule for the Clifford algebra $C(V^*\oplus V)$.

Let us assume that the bilinear form $(,)$ on $V$ is non-degenerate.
Then the linear map
$$
              V\to V^*,\ v\mapsto v^\#,\ \qquad v^\#(v')=(v,v')
$$
establishes a natural isomorphism between the vector space $V$ and its dual.
In this situation we need no longer distinguish between the two spaces,
$V$ and $V^*$, or distinguish between $v$ and $v^\#$, i.e., we shall write
$\io(v)$ where we really mean $\io(v^\#)$.

Consider the following Clifford map:
$$
        c:V\to{\rm End}\,\bw V,\qquad c(v)=\ep(v)-\io(v).
$$
Its extension $c:C(V)\to{\rm End}\,\bw V$ gives the exterior algebra
$\bw V$ the structure of a selfadjoint $C(V)$ supermodul. It is always 
reducible. To prove selfadjointness we must first extend $(,)$ on $V$ to a
bilinear form on $\bw V$. This is done in a standard fashion:
\begin{eqnarray*}
   (1,1) &=&1\\ 
   (\bw^k V,\bw^l V)&=&0\qquad (k\ne l)\\
  (v_1\wedge\ldots\wedge v_k,v'_1\wedge\ldots\wedge v'_k)
         &=&\det(v_i,v'_j)_{i,j=1}^k\qquad (k=1,\ldots,n).
\end{eqnarray*}
Laplace's expansion formula for determinants is simply stated as
$\ep(v)^*=\io(v)$. Therefore, $\io(v)^*=\ep(v)$ and
$$
     c(v)^*=\Big(\epsilon(v)-\iota(v)\Big)^*=-c(v)\qquad(v\in V)
$$
which suffices to establish selfadjointness of the Clifford modul.

It helps the physical intuition to compare the formalism with that of
the CAR algebra used in the theory of fermions. For it is clear that
the Clifford action on $\bw V$, the `Fock space' in physics, is
fully determined by the conditions:
\begin{eqnarray*}
     \{\io(v_1),\ep(v_2)\}   &=& (v_1,v_2),\qquad \io(v)1=0\\
     \{\io(v_1),\io(v_2)\}&=& 0\ =\ \{\ep(v_1),\ep(v_2)\}.
\end{eqnarray*}
Thus, $\io(v)$ and $\ep(v)$ may be viewed as annihilation and
creation operators respectively, while the unit $1\in\bw V$ serves as
the `vacuum'.

The $C(V)$ module structure of $\bw V$ now allows us to set up an isomorphism
between vector spaces, 
$$
        \sigma:C(V)\to\bw V,\qquad \sigma(a)=c(a)1,
$$
known as the {\em symbol map}.
Since $c$ (and hence $\sigma$) respects the $\ZZ_2$-grading, $\sigma$ is
in fact an isomorphism between superspaces though not an algebraic
isomorphism. For example, with $v_i\in V$ we have
\begin{eqnarray*}
  \sigma(1) &=& 1\\
  \sigma(v_1) &=& v_1\\
  \sigma(v_1v_2)&=& v_1\wedge v_2-(v_1,v_2)1\\
  \sigma(v_1v_2v_3)&=&v_1\wedge v_2\wedge v_3
                    -(v_1,v_2)v_3+(v_3,v_1)v_2-(v_2,v_3)v_1.
\end{eqnarray*}
If $e_i$ is a basis in $V$, assuming that $(e_i,e_k)=0$ for $i\ne k$, 
let $e_I$ be the induced basis in $C(V)$ given by Eq.(\ref{eq:bas}). 
Then the calculation (using an inductive argument and the fact that 
contractions are absent)
\begin{eqnarray*}
\sigma(e_I)=\sigma(e_{i_1}\ldots e_{i_k})
   &=&c(e_{i_1}\ldots e_{i_k})1\\
   &=&c(e_{i_1})\cdots c(e_{i_k})1\\
   &=&c(e_{i_1})\cdots c(e_{i_{k-1}})e_{i_k}\\
   &=&c(e_{i_1})\cdots c(e_{i_{k-2}})(e_{i_{k-1}}\wedge e_{i_k})\\
   &=&\ldots=e_{i_1}\wedge\ldots\wedge e_{i_k}\,
\end{eqnarray*}
reveals that the basis in $C(V)$ is mapped onto the corresponding basis
in $\bw V$. 

A frequently used map is the inverse $\sigma^{-1}:\bw V\to C(V)$. It is
referred to as the {\em quantization map} because, intuitively,
one likes to think of the Clifford algebra $C(V)$ as a quantum deformation
of the `classical' (supercommutative) algebra $\bw V$. 
The quantum deformation is visible in contractions terms involving 
$(v_i,v_k)$ and disappears when $(,)$ vanishes identically on $V$, 
i.e., when $\sigma:\bw V\to \bw V$ reduces to the identity map.

It may be instructive to see the quantum analogues of $k$-vectors:
$$
      \sigma^{-1}(v_1\wedge\cdots\wedge v_k) = \frac{1}{k!}\sum_\pi
                  \mbox{sign}(\pi)\,v_{\pi(1)}v_{\pi(2)}\cdots v_{\pi(k)}\ .
$$
In principle, the use of permutations $\pi$ can be avoided here and one may write
instead:
\begin{eqnarray*}
      \sigma^{-1}(1) &=&1\\
      \sigma^{-1}(v_1) &=&v_1\\
      \sigma^{-1}(v_1\wedge v_2)&=& v_1v_2+(v_1,v_2)1\\
      \sigma^{-1}(v_1\wedge v_2\wedge v_3) &=& v_1v_2v_3
                  +(v_1,v_2)v_3-(v_3,v_1)v_2+(v_2,v_3)v_1.
\end{eqnarray*}
The quantization map can now be used to carry the $\ZZ$-grading of the exterior
algebra $\bw V=\sum_k\bw^kV$ to the Clifford algebra $C(V)$:
$$\textstyle
   C(V)=\sum_kC^k(V)\quad\mbox{where}\quad C^k(V):=\sigma^{-1}(\bw^kV)\,.
$$
Of course, the sum has only $n+1$ terms since $C^k(V)=\{0\}$ if either
$k<0$ or $k>n$. Note that the $\ZZ$-graded vector space $C(V)$ does not
satisfy the conditions of $\ZZ$-graded algebra since the quantization map
$\sigma^{-1}$ fails to be an algebraic isomorphism. For instance,
$$
     v_1v_2+(v_1,v_2)1\in C^2(V)\qquad(v_i\in V)
$$
and hence the upper index $k$ in $C^k(V)$ should be interpreted and used
with great care.
Nevertheless, the $\ZZ$-grading (as vector space) is consistent with
the $\ZZ_2$-grading (as algebra):
$$
        C^+(V)=\sum_{k={\rm even}}C^k(V),\qquad
        C^-(V)=\sum_{k={\rm odd}}C^k(V).
$$
Note in particular that $C^0(V)=\RR$ and $C^1(V)=V$. Also,
$$
   \dim C^k(V)=\dim \bw^kV ={n\choose k}.
$$
In applications it is important to realize that the map $a\mapsto a^*$,
that is taking adjoints $C(V)$, respects the $\ZZ$-grading. In fact,
for a basis $e_I$ in $C^k(V)$ with $|I|=k$, we get
$$
    e_I^*=(e_{i_1}\cdots e_{i_k})^*=(-1)^ke_{i_k}\cdots e_{i_1}
    =(-1)^k(-1)^{k(k-1)/2}e_I =(-1)^{k(k+1)/2}e_I
$$
and thus $a^*=(-1)^{k(k+1)/2}a$ for all $a\in C^k(V)$.

\section{The Spin Group}

In dealing with a superalgebra $A$ the notion of the {\em supercommutator\/}
of two of its elements will be important:
$$
     \sco{a,b}=\cases{ab+ba & if $a,b\in A^-$\cr ab-ba &otherwise.\cr}
$$
Supercommutators $\sco{a,.}$ with respect to some fixed element $a\in A$
may then be viewed as {\em inner derivations\/} of the algebra $A$ owing to
the formula
\begin{equation}
  \label{eq:ind}
   \sco{a,bc}=\cases{\sco{a,b}c-b\sco{a,c} & if $a,b\in A^-$\cr
                     \sco{a,b}c+b\sco{a,c} & otherwise.\cr}
\end{equation}
There are two remarkable properties of the subspaces $C^k(V)\subset C(V)$:
\begin{eqnarray}
          \sco{C^1(V),C^k(V)}&\subset& C^{k-1}(V)   \label{c1}\\
          \sco{C^2(V),C^k(V)}&\subset& C^k(V)\,.    \label{c2}
\end{eqnarray}
These inclusions are consequences of the fact that, for any $v\in V$, the
following diagram commutes:
$$
\def\normalbaslines{\baselineskip30pt\lineskip5pt\lineskiplimit5pt}
\matrix{C(V)&\mapr{\sco{v,\cdot}}&C(V)\cr
        \mapd{\sigma}&  &\mapd{\sigma}\cr
        \bw V & \mapr{-2\io(v)} &\bw V\cr}
$$
The isomorphism $\sigma$ makes the supercommutator and the contraction 
correspond, or phrased differently, the formula
\begin{equation}
  \label{eq:sis}
  -\hf\sco{v,a}=\sigma^{-1}\Big(\io(v)\sigma(a)\Big)\qquad(v\in V,a\in C(V))
\end{equation}
extends the Clifford relation $\sco{v,v'}=-2(v,v')1$. To prove  
Eq.(\ref{eq:sis}) it suffices to show its validity for a basis in $C(V)$. 
Let $e_I$ be the basis elements given by Eq.(\ref{eq:bas}). Setting $a=e_I$ 
with $|I|=k$ we obtain:
\begin{eqnarray*} 
 \sco{v,e_I} &=&\sco{v,e_{i_1}\cdots e_{i_k}}\\
             &=&
\sco{v,e_{i_1}} e_{i_2}\cdots e_{i_k}-e_{i_1}\sco{v,e_{i_2}\cdots e_{i_k}}\\
                             &=&
   -2(v,e_{i_1})e_{i_2}\cdots e_{i_k}-e_{i_1}\sco{v,e_{i_2}\ldots e_{i_k}}\\
                             &=&\ldots=
-2\sum_{l=1}^k(-1)^{l+1}(v,e_{i_l})e_{i_1}\ldots \hat{e}_{i_l}\cdots e_{i_k}\\
                             &=&
-2\sigma^{-1}\Big(\io(v)(e_{i_1}\wedge\ldots\wedge e_{i_k})\Big) 
  =-2\sigma^{-1}\Big(\io(v)\sigma(e_I)\Big). 
\end{eqnarray*}
Having proved Eq.(\ref{eq:sis}) we see that the statement (\ref{c1}) is an 
easy consequence, simply because the contraction operator $\io(v)$ has degree
$-1$, that is, it maps
$\bw^kV$ into $\bw^{k-1}V$. The second statement (\ref{c2}) is more involved.
Working with a basis, we first find that, for $i\ne j$,
$$
\sco{e_ie_j,e_I}=e_i\sco{e_j,e_I} 
                  +(-1)^{|I|} \sco{e_i,e_I}e_j.
$$
The first term on the right hand side describes a replacement of $j\in I$
by $i$, the second of $i\in I$ by $j$. The result, of course, could also
be zero: for the first term if $i\in I$ or $j\nein I$, for the second term
if $j\in I$ or $i\nein I$. In the case $i,j\in I$ one uses
$\sco{e_ie_j,e_ie_j}=0$ to show that $\sco{e_ie_j,e_I}=0$.
In any case, the result is in $C^k(V)$.

Eq.(\ref{c2}) plays a decisive role in a number of applications, because
the supercommutator, if suitably restricted, preserves the $\ZZ$-grading.
Note that, in the case of Eq.(\ref{c2}), the supercommutator is in fact
the commutator. Our interest lies in the case $k=2$,
$$ 
    [ \:\:,\:\, ]: C^2(V)\times C^2(V)\to C^2(V),
$$
but also in the case $k=1$:
$$
         [ C^2(V),V ] \subset V\,.
$$ 
Thus, $C^2(V)$ is a Lie algebra operating on $V$ via the adjoint representation:
\begin{eqnarray*}
             C^2(V)&\arr{\sad} &\End V\\
             a     &\mapsto   &[a,\cdot\,]\ .
\end{eqnarray*}
Obviously, $\ad[a,b]=[\ad a,\ad b]$. Note also the relation
\begin{equation}
  \label{so}
  (\ad(a)v,v')+(v,\ad(a)v')=0\qquad (v,v'\in V)
\end{equation}
which states that $\ad(a)$ is an element of $so(V)$, the Lie algebra of
$SO(V)$. Since
$$
    \dim so(V)=\dim C^2(V) = {n\choose 2},\qquad \ker\ad={0},
$$
the map $\ad:C^2(V)\to so(V)$ turns out to be a Lie isomorphism. It remains
to prove Eq.(\ref{so}). The argument runs as follows:
\begin{eqnarray*}
(\ad(a)v,v')+(v,\ad(a)v')&=&-\hf(\sco{\sco{a,v},v'}+\sco{v,\sco{a,v'}}) \\ 
                         &=&-\hf\sco{a,\sco{v,v'}} \\
                         &=&\sco{a,(v,v')1}\ =\ 0.                             
\end{eqnarray*}
In the first and the third step, we used the Clifford relation while in the
second step we applied the generalized Jacobi identity,
\begin{equation}
  \label{jac}
\sco{a,\sco{b,c}}-\sco{\sco{a,b},c}=\cases{-\sco{b,\sco{a,c}}& 
                                           if $a,b\in A^-$\cr
                                 \phantom{-}\sco{b,\sco{a,c}}&
                                                otherwise,\cr}
\end{equation}
valid in any superalgebra $A$.

To pass from a Lie algebra of operators to a Lie group is a straightforward 
procedure known as the exponential mapping. The resulting group
$$ 
            \Spin(V)=\exp C^2(V)\subset C^+(V)
$$
is called the {\em spin group}. By the above construction, this group 
is seen to act on the vector space $V$ via the adjoint representation
$\Ad=\exp\circ\,\ad$. Hausdorff's formula then provides a more explicit
description:
$$
      \Ad(e^a)v=e^{\sad(a)}v=e^a\,v\,e^{-a}\,.
$$ 
Notice that the product on the right hand side has to be taken
within $C(V)$. The above analysis guarantees that the result will again
be in $V\subset C(V)$.

If $V$ is Euclidean and $\dim V>1$ or if the bilinear form $(,)$ is
nondegenerate and $\dim V>2$, then, setting $\ZZ_2=\{1,-1\}$, the 
following diagram has exact sequences as its rows:
$$
\def\normalbaslines{\baselineskip30pt\lineskip5pt\lineskiplimit5pt}\matrix{%
 &         &0&\mapr{\ }&C^2(V)      &\mapr{\sad}&so(V)      &\mapr{\ }0\cr
 &         & &         &\mapd{\exp} &           &\mapd{\exp}&          \cr  
1&\mapr{\ }&\ZZ_2&\mapr{\ }&\Spin(V)&\mapr{\sAd} &SO(V)      &\mapr{\ }1\cr}
$$
Typically, the spin group is a double covering of $SO(V)$. For the
convenience of the reader, we include the proof. The exactness of the
top sequence has already been demonstrated. As for the bottom sequence,
we need only show that $\ker\Ad=\ZZ_2$. This is done in three steps.
\\[3mm]
(1) If $g\in\ker\Ad$ (i.e., $\Ad(g)=1$), then $gvg^{-1}=v$ for all $v\in V$
or $\sco{v,g}=0$ since $g\in C^+$. From Eq.(\ref{eq:sis}) we infer that
$\io(v)\sigma(g)=0$ implying $\sigma(g)\in\bw^0V$ or $g\in C^0(V)\cong\RR$.
\\[3mm]
(2) Under the above assumptions on $V$ we may find two vectors, $v$ and $w$,
such that $(v,w)=0$ and $(v,v)=(w,w)=\pm 1$. Since $vw\in C^2(V)$ and
$(vw)^2=vwvw=-vvww=-(v,v)(w,w)=-1$,
$$
      \exp(tvw)=(\cos t)1+(\sin t)vw\in\Spin(V)\qquad (t\in\RR).
$$
In particular, for $t=\pi$, we learn that $-1\in\Spin(V)$ and hence
$\ZZ_2\subset\Spin(V)$. Obviously, $\Ad(-1)=1$ and thus $\ZZ_2\subset
\ker\Ad$.
\\[3mm]
(3) To demonstrate the equality $\ZZ_2=\ker\Ad$ we recall that the map 
$a\mapsto a^*$ leaves $C^k(V)$ invariant and $a^*=-a$ for $a\in C^2(V)$.
Thus, any group element $g=\exp(a)\in\Spin(V)$ satisfies $gg^*=1$, because
$$
      g^* = \exp(a^*)=\exp(-a)=g^{-1},
$$
and moreover, if $g\in C^0(V)$, then also $g^2=1$ and hence $g\in\{1,-1\}$ 
which completes the proof.
\\[3mm]
If $E$ is a (real) Clifford module, the Clifford action $c$ automatically 
induces a representation $c$ of the spin group on $E$. If the module is
selfadjoint, then $c(g)^*=c(g^*)$ and the representation is orthogonal in the 
sense that $c(g)^*=c(g)^{-1}$, i.e., $c(g)\in SO(E)$. If $E$ is a supermodule,
the representation is reducible: the subspaces $E^\pm$ turn out to be invariant
since $\Spin(V)\subset C^+(V)$.

To illustrate the foregoing discussion we will study two examples relevant
for physics.

{\bf Example 1}. Let $V$ be the Euclidean space $E_3$. With respect to
some (orthonormal) basis $(e_i)_{i=1}^3$ in $E_3$, the Clifford algebra 
$C(E_3)$ of dimension 8 is defined through the relations 
$e_ie_j+e_je_i+2\delta_{ij}1=0$.
The Lie algebra $C^2(E_3)$ is 3-dimensional with basis
$a_i:=\frac{1}{4}\epsilon_{ijk}e_je_k$ $(i=1,2,3)$. The commutator relations
\[
         [a_i,a_j]=\ep_{ijk}a_k
\]
are those of the Lie algebra $su(2)$. We may prove now that the adjoint action
on $E_3$ is given by
$$
      \ad(a_i)=A_i,\qquad (A_i)_{jk}:=-\ep_{ijk}.
$$
Indeed, for $v\in E_3$,
\begin{eqnarray*}
\ad(a_i)v&=&\sco{a_i,v}=\frc{1}{4}\ep_{ijk}\sco{e_je_k,v}\\
         &=&\frc{1}{4}\ep_{ijk}\Big(e_j\sco{e_k,v}-\sco{e_j,v}e_k\Big)\\
         &=&\frac{1}{4}\ep_{ijk}\Big(-2e_j (e_k,v)+2(e_j,v) e_k\Big)\\
         &=&-\ep_{ijk}(e_k,v)e_j=(A_i)_{jk}(e_k,v) e_j\ =\ A_iv
\end{eqnarray*}
and hence the adjoint representation of the group $\Spin(E_3)$ on $E_3$ given 
by
$$
    \Ad(e^a)v=e^Av, \qquad v\in E_3,\ a\in C^2(E_3),\ A=\ad(a)
$$ 
is but the familiar action of the group $SO(3)$. The spin group itself, 
a double cover of $SO(3)$, is thus seen to be isomorphic to the unitary 
group $SU(2)$.
\vspace{2mm}\par
{\bf Example 2}. Let $V$ be the 4-dimensional Minkowski space $M_4$ with
metric $g_{\mu\nu}=\mbox{diag}(1,-1,-1,-1)$ $(\mu,\nu=0,\ldots,3)$ 
and standard basis vectors $e_\mu$. The Clifford algebra $C(M_4)$ of dimension
16 is defined through the relations  $e_\mu e_\nu+e_\nu e_\mu+2g_{\mu\nu}1=0$.
The Lie algebra $C^2(M_4)$ is 6-dimensional with basis 
$m_{\mu\nu}:=-\frac{1}{2}e_\mu e_\nu$ ($\mu<\nu$). It is convenient to regard
$m_{\mu\nu}$ as an antisymmetric tensor. The commutation relations
$$
  [m_{\mu\nu},m_{\sigma\tau}]=g_{\tau\mu}m_{\nu\sigma}+g_{\mu\sigma}m_{\tau\nu}
                             +g_{\tau\nu}m_{\sigma\mu}+g_{\nu\sigma}m_{\mu\tau}
$$
are those of the Lie algebra $sl(2,\CC)$, and the adjoint action on $M_4$ 
is given by
$$
   \ad(m_{\mu\nu})=M_{\mu\nu},\qquad 
   (M_{\mu\nu})_{\alpha\beta}=
    g_{\mu\alpha}g_{\nu\beta}-g_{\mu\beta}g_{\nu\alpha}
$$
implying that the adjoint action $\Ad=\exp\circ\,\ad$ of the spin group
$\Spin(M_4)\cong SL(2,\CC)$ on $M_4$ coincides with the action of the 
Lorentz group (the identity component thereof, strictly speaking). 
As is well known, the group $SL(2,\CC)$ is a double covering of the
Lorentz group.

We cannot, however, incorporate Dirac's $\gamma$ matrices in the present
framework unless we are willing to complexify the Clifford algebra and
to study complex Clifford modules. For it is clear that, with regard to
the last example, $\gamma_\mu$ corresponds to $ie_\mu$, where $i$ is 
the imaginary unit,
and the spinor space, on which the $\gamma$'s act, is isomorphic to $\CC^4$.
Hence, with respect to some basis in $\CC^4$, the $\gamma$'s are
represented by complex matrices. The necessary steps to deal with this
problem of complex extension are the subject of the next sections.

\section{The Spinor Module}

When dealing with the real vector space $V$ of dimension $n$,                  
the complexified Clifford algebra of (complex) dimension $2^n$ is 
the tensor product $C(V)\otimes\CC$. The concept of a {\em complex Clifford
module\/} is then obvious: it is some complex vector space $E$ together with
an action of $C(V)\otimes\CC$ respecting the complex linear structure.
The goal now is to construct, for suitable $V$'s, a canonical irreducible 
complex Clifford module, $S$, termed the {\em spinor module}, i.e., we want
to construct an isomorphism
\begin{equation}
  \label{ccv}
               c:C(V)\otimes\CC\to\End S.
\end{equation}
This gives $\End S$ the dimension $2^n$ and consequently $S$ the dimension 
$2^{n/2}$ which makes sense provided $n=\,$ even. In Minkowski's model of 
spacetime, where $V=M_4$, we are fortunate to encounter an even
dimension, $n=4$, and so $\dim S=2^2=4$ which is the dimension of Dirac
spinors, the elements of the spin module.

To construct the spinor module in general we pass to the complex space 
$V\otimes\CC$ first and then extend the quadratic form $q(v)$ in $V$ to a 
quadratic form $q(w)$ in $V\otimes\CC$. Such an extension is unique.
Moreover, there is a  complex-bilinear form $(,)$ in $V\otimes\CC$ 
extending the real-bilinear form in $V$ and satisfying $q(w)=(w,w)$. 
Supposing there are complimentary subspaces $V_\pm$ of $V\otimes\CC$, dual to
each other in the sense that $V_\pm^*\cong V_\mp$, then the isomorphism
$$
             V\otimes\CC\cong V^*_+\oplus V_+
$$
is said to provide a complex {\em polarisation\/} of $V$ if the quadratic
form $q$ in $V\otimes\CC$ corresponds to the canonical quadratic form in
$V^*_+\oplus V_+$, that is to say, if 
$$
            \hf q(w)=w^\#_-(w_+)=(w_-,w_+)
$$ 
for $w_\pm\in V_\pm$ such that $w=w_++w_-$. The factor $\hf$ in inevitable
because $(w_\pm,w_\pm)=0$ and $q(w)=(w_++w_-,w_++w_-)=2(w_-,w_+)$.
\vspace{2mm}\par\noindent
{\bf Example 1}. The four-dimensional Minkowski space, $M_4$, admits a
polarisation because the Lorentz metric, extended to $M_4\otimes\CC$,
may be written in a polarized form:
$$
        (x,x) =(x^1+ix^2)(-x^1+ix^2)+(x^0+x^3)(x^0-x^3) \qquad (x^\mu\in\CC).
$$
Thus, a possible choice of the subspaces is
$$
   V_\pm =\{x\in M_4\otimes\CC\ |\ \pm x_1+ix_2=x^0\pm x^3=0\}\cong\CC^2.
$$
{\bf Example 2.}. The four-dimensional Euclidean space, $E_4$, admits a
polarisation because the Euclidean metric, extended to $E_4\otimes\CC$,
may be written in a polarized form:
$$
    (x,x) =(x^1+ix^2)(x^1-ix^2)+(x^3+ix^4)(x^3-ix^4) \qquad (x^\mu\in\CC).
$$
A possible choice of the subspaces is
$$
   V_\pm =\{x\in E_4\otimes\CC\ |\  x^1\pm ix^2=x^3\pm ix^4=0\}\cong\CC^2.
$$

The two previous examples are typical in that they demonstrate a general fact:
any real pseudo-Euclidean space of even dimension admits a complex 
polarisation. The space $V$ is said to be {\em pseudo-Euclidean\/} if there
is a bilinear form preserving isomorphism
$$
                V\otimes\CC\cong E_n\otimes\CC
$$
where $E_n$ denotes the $n$-dimensional Euclidean space. Equivalently stated, 
one can find a basis $e_j$ in $V\otimes\CC$ such that
$$
         (e_j,e_k)=\delta_{jk}\qquad (j,k=1,\ldots,n).
$$
Such a basis will be called {\em orthonormal}. If $V$ happens to be Euclidean,
any orthonormal basis in $V$ would do. 

Note, there are many cases falling into the category of pseudo-Euclidean 
spaces, such as the Minkowski space $M_4$, the deSitter (dS), and the
Anti-deSitter (AdS) space.
However, with the present emphasis on bilinear forms rather than on
scalar products, the distinction between Euclidean and pseudo-Euclidean
spaces disappears in the complex domain. From now on we shall always assume
that $V$ is (pseudo-)Euclidean and even-dimensional.

It is not difficult to demonstrate that $V$ admits a polarisation. The reason 
is this. Given an orthonormal basis $(e_i)_{i=1}^n$ in $V\otimes\CC$, we can 
define complementary subspaces by
$$
     V_\pm = \mbox{span}\Big\{e_{2k-1}\pm ie_{2k}\,|\,k=1,\ldots,n/2\Big\}
$$
and set up isomorphisms $V_\pm\to V_\mp^*$, $w\mapsto w^\#$, by
$$
           w^\#(w')=(w,w'),\qquad  w'\in V_\mp\,.
$$
It remains to prove that
\begin{equation}
  \label{qw}
          (w_-,w_+)=\hf q(w)\qquad (w=w_++w_-,\ w_\pm\in V_\pm).  
\end{equation}
To this end we write $w= x^ke_k\in V\otimes\CC$ with coordinates 
$x^k\in\CC$ so that $q(w)=\sum_k (x^k)^2$ and $w=w_++w_-$ with
$$
         w_\pm=\hf\sum_{k=1}^{n/2}(x^{2k-1}\mp ix^{2k})(e_{2k-1}\pm ie_{2k}).
$$
Then
\begin{eqnarray*}
   (w_-,w_+) &=& \hf\sum(x^{2k-1}+ix^{2k})(x^{2k-1}-ix^{2k})\\
             &=& \hf\sum (x^k)^2 = \hf q(w)\,.
\end{eqnarray*}
Consider now the exterior algebra
$$
                 S=\bw V_+\ .
$$
As complex vector space, $S$ has the dimension $2^{n/2}$. To give $S$ the 
structure of a Clifford module, we need only know the action of $v\in V$ 
on $S$. We define $c: V\to\End S$ by
$$
  c(v)=\sqrt{2}\Big(\ep(w_+)-\io(w_-)\Big),
  \qquad v=w_+ + w_-,\qquad w_\pm\in V_\pm
$$
and immediately verify that $c$ is a Clifford map:
$$
  c(v)^2=-2\sco{\ep(w_+),\io(w_-)}=-2(w_+,w_-)=-q(v).
$$
Its extension to $V\otimes\CC$ and $C(V)\otimes\CC$ is straightforward.
As is well known and will be shown later, the Clifford algebra is {\em simple}.
Irreducibitity of a Clifford module $E$ then is equivalent to 
the map $C(V)\otimes\CC\cong\End E$ being an isomorphism. The proof that the 
spinor module $S$ is irreducible offers no problem and will be skipped.

We now show that the spinor module $S$ is in fact a supermodule. Namely, 
with respect to an orthonormal basis in $V\otimes\CC$ we define the 
{\em chirality operator\/} $\Gamma\in C(V)\otimes\CC$ by
$$
            \Gamma = i^{n/2}e_1e_2\cdots e_n.
$$
The factor in front has been chosen so as to guarantee that $\Gamma^2=1$.
In physics where $V=M_4$, the corresponding operator $c(\Gamma)$ on $S$ is 
known as the $\gamma_5$ matrix.

The question arises whether the definition of $\Gamma$
depends on our choice of the basis or `frame'. Suppose that $e'_j$ is another
orthogonal basis in $V\otimes\CC$. Then $e'_j=A_{jk}e_k$ for some complex
matrix $A$ with $A^TA=1$ so as to preserve orthogonality of the basis.
Consequently, $\det A=\pm 1$. It is clear now that there
are precisely two classes of frames or {\em orientations},
the two frames $(e_k)$ and $(e'_k)$ have the same orientation if $\det A=1$.

From the definition of the quantization map we infer that
$$
      e_1'\cdots e_n' =\sigma^{-1}(e_1'\wedge\cdots\wedge e_n')
     = \det A\ \sigma^{-1}(e_1\wedge\cdots\wedge e_n)
     = \pm  e_1\cdots e_n
$$
which tells us that, given some orientation, there will be no ambiguity 
in the definition of $\Gamma$ because then $\det A=1$ always. For the sake of 
consistency, we define both $S$ and $\Gamma$ with 
respect to some orthogonal and oriented basis in $V$.

Since $n$ is even, $\Gamma a=\pm a\Gamma$ for $a\in C^\pm(V)\otimes\CC$.
Therefore, if we define the projection operators
$$
              p^\pm =\hf(1\pm c(\Gamma))\in \End S\ ,
$$ 
we get a grading of the spinor module respected by the Clifford action:
$$
            S=S^+\oplus S^-,\qquad S^\pm = p^\pm S\ .
$$
There seem to be two different $\ZZ_2$-gradings on $S$, both of them
respected by the Clifford action: one given by the chirality operator 
and another one given by the structure of $S$ an exterior
algebra: $S^\pm=\bw^\pm V_+$. It is important to realize that these two
gradings coincide. The proof is facilitated by choosing a suitable basis,
$$
    e_k^\pm=\frc{1}{\sqrt{2}}(e_{2k-1}\pm ie_{2k})\in V_\pm\ ,
$$
and related multiplication and contraction operators,
$$
  \io_k=\io(e^-_k),\qquad\ep_k=\ep(e^+_k),
$$
so that 
$$
     \sco{\ep_k,\io_l}=(e^-_k,e^+_l)=\delta_{kl}.
$$
We introduce operators $a_k$ which commute:
$$
    a_k := \hf (e_k^+e_k^- - e_k^-e_k^+)=ie_{2k-1}e_{2k}\in C^+(V)\otimes\CC.
$$
It is immediately clear that the chirality operator assumes the form
\begin{equation}
  \label{ak}
      \Gamma=a_1a_2\cdots a_{n/2}
\end{equation}
while $c(a_k)$ may be written in terms of multiplication and contraction 
operators:
$$
  c(a_k) = \io_k\ep_k-\ep_k\io_k  = 1-2\ep_k\io_k\,.
$$
For any $k$, the product $\ep_k\io_k$ has eigenvalues $0,1$ and, therefore,
$c(a_k)$ has eigenvalues $\pm 1$. Namely, on the basis
$$
      e_I=e^+_{i_1}\wedge\ldots\wedge e^+_{i_p}\in \bw^p S
$$
we obtain
$$
  c(a_k)e_I=\cases{\phantom{-}e_I & if $k\nein I$\cr
                             -e_I & if $k\in   I$\cr}
$$
and hence $c(\Gamma)e_I=(-1)^{|I|}e_I$ which completes the proof.

In passing we remark that the operators $a_k$ are elements of the
complex Lie algebra $C^2(V)\otimes\CC$. Moreover,
$$
             ia_k\in\Spinc V=\exp\Big(C^2(V)\otimes\CC\Big)
$$
as can be seen from $a^2_k=1$ and
$$
        \exp\left(i\frac{\pi}{2}a_k\right)=ia_k\ .
$$
By Eq.(\ref{ak}), the chirality operator $\Gamma$ is in $\Spinc V$.
In the Euclidean case, choosing a basis $e_j\in E_n$, we have
$ia_k\in C^2(E_n)$, but also $ia_k\in \Spin E_n$. By Eq.(\ref{ak}),
$\Gamma\in\Spin E_n$ iff $i^{n/2}\in\RR$, i.e., iff $n=0\!\!\!\pmod 4$.

\section{Selfadjointness for Complex Modules}

We want to extend the concept of selfadjointness to complex 
modules when there is an isomorphism $V\otimes\CC\cong E_n\otimes\CC$. 
We may thus take any orthogonal basis in $E_n$ and regard it as orthogonal 
basis in $V\otimes\CC$. The implied polarisation has the property
$$
                \overline{V_\pm} = V_\mp
$$
where $w\mapsto\bar{w}$ means {\em complex conjugation\/} in $E_n\otimes\CC$.
To the previously listed
properties, satisfied by the \Star operation, we have to add only its
antilinearity, $(\lambda a)^*=\bar{\lambda}a^*$. Then the standard \Star 
operation in the real Clifford algebra $C(V)$ has a unique extension to its 
complex counterpart, the algebra $C(V)\otimes\CC$. In particular, 
$$
       w^*=-\bar{w},\qquad w\in V\otimes\CC\ . 
$$
If $E$ is some complex Clifford module, then in order to give $\End E$
the structure of a \Star algebra we need to have a Hermitian  
structure on $E$. For, if $\br{,}$ is a scalar product in $E$,
then the adjoint of $A\in\End E$ is given by
$$
              \br{A^*x,y} =\br{x,Ay}\qquad (x,y\in E).
$$
A complex Clifford module $E$ with Clifford action $c:C(V)\otimes\CC\to\End E$
satifying
$$
      c(a^*)=c(a)^*
$$
is said to be {\em selfadjoint}. In particular, the relation
$$
                    c(w)^*+c(\bar{w})=0
$$
holds which, as we know, suffices to establish selfadjointness. 

To demonstrate that the spinor module $S$ is selfajoint, we need to specify
a scalar product in $V\otimes\CC$,
$$
          \br{w,w'}=(\bar{w},w'),
$$
restrict it to $V_+$, subspace of $V\otimes\CC$, and then to extend $\br{,}$ 
to all of $\bw V_+$:
\begin{eqnarray*}
   \br{1,1} &=&1\\ 
  \br{\bw^k V_+,\bw^l V_+}&=&0\qquad (k\ne l)\\
  (w_1\wedge\ldots\wedge w_k,w'_1\wedge\ldots\wedge w'_k)
         &=&\det\br{w_i,w'_j}_{i,j=1}^k\qquad (k=1,\ldots,n/2).
\end{eqnarray*}
It is easy to see that $\br{,}$ is indeed a scalar product on $V\otimes\CC$\,: 
if $w=x^ke_k$, then $\bar{w}=\bar{x}^ke_k$ and 
$\br{w,w}=(\bar{w},w)=\sum |x^k|^2$. Moreover, $V_+$ and $V_-$ are
orthogonal subspaces. 

To summarize, we have passed from the bilinear form $(,)$ in $V\otimes\CC$ to
the scalar product $\br{,}$. It seems natural to change the definition of
the contraction operator accordingly,
$$
              \bio(w)=\io(\bar{w})\qquad (w\in V_+),
$$
so that $\bio(w)(w'\wedge a)=\br{w,w'}a -w'\wedge\bio(w)a$ in $\bw V_+$.
Under the \Star operation in $\End\bw V_+$ the
behavior of the multiplication and contraction operators (which generate
the endomorphism algebra) is as follows:
$$
   \ep(w)^*=\bio(w),\qquad \bio(w)^*=\ep(w)\qquad(w\in V_+).
$$
Consequently, for $w=w_1+\bar{w}_2$ with $w_j\in V_+$ (such decomposition
is unique), 
\begin{eqnarray*}
 c(w)^* &=&\sqrt{2}(\ep(w_1)-\bio(w_2))^*\\
        &=&\sqrt{2}(\bio(w_1)-\ep(w_2)) =-c(\bar{w})=c(w^*)
\end{eqnarray*}
thereby proving that the spinor module $S$ is selfadjoint.

The case of a pseudo-Euclidean vector space $V$ is in no way different from
the case of a Euclidean vector space, except that `complex conjugation' 
receives a different meaning: it is defined in $E_n\otimes\CC$ rather
than in $V\otimes\CC$. Therefore, $v\in V$ is not a `real' element of 
$E_4\otimes\CC$ unless $V\cong E_n$. As a consequence, we
loose the property $c(v)^*+c(v)=0$. This observation may be phrased as
follows: as $C(V)$ modul, the spinor module $S$ {\em is not selfadjoint\/}
unless $V$ is Euclidean. So the lesson is: complex modules ought to be
regarded as $C(V)\otimes\CC$ modules.

{\bf Example}. In Dirac's relativistic theory of the electron, the Hilbert
space is $L^2(\RR^3)\otimes S$ where $S$ is the spinor module with
scalar product $\br{,}$ as constructed above. The Clifford algebra is
$C(M^*_4)\otimes\CC$ with $M^*_4$ the so-called {\em momentum space}. It 
is dual to the Minkowski space $M_4$. If $e_\mu$ is the standard basis in
$M_4$, we let $e^\mu$ denote the dual basis in $M^*_4$. Hence, any 
$p\in M^*_4$ is of the form $p=p_\mu e^\mu$ with $p_\mu\in\RR$.
To make effective use of the isomorphism 
$$
           M^*_4\otimes\CC\cong E_4\otimes\CC
$$
we need to introduce complex momenta, too. There is an orthogonal basis
in $E_4$ which, under the above isomorphism, corresponds to the vectors
$$
             e^0,\ ie^1,\ ie^2,\ ie^3\ 
$$ 
and the preferred way to expand complex momenta is:
$$
   p = p_0 e^0+p_1ie^1+p_2ie^2+p_3ie^3 \qquad(p_\mu\in\CC).
$$ 
If $p$ has complex coordinates $p_\mu$, then $\bar{p}$ has the complex
conjugate coordinates $\bar{p}^\mu$. Another way of stating this peculiar 
property is:
$$
        \bar{e}^0=e^0,\ \bar{e}^k=-e^k,\ k=1,2,3.
$$
Restricted to $M^*_4$, the operation $p\mapsto\bar{p}$ is thus seen to
coincide with the reflection in 3-space (or parity operation). The
chirality operator is
$$
     \Gamma =i^2e^0(ie^1)(ie^2)(ie^3)=ie^0e^1e^2e^3\ \in C(M^*_4)\otimes\CC.
$$
To make contact with Dirac's theory, we define
$$
          \gamma^\mu =ic(e^\mu),\qquad \gamma_5=c(\Gamma)
$$
so that $\gamma_5=i\gamma^0\gamma^1\gamma^2\gamma^3$. The \Star operation
sending $p$ to $p^*=-\bar{p}$ acts on the basis vectors as follows:
$$
         e^{0*}=-e^0,\ e^{k*}=e^k,\ k=1,2,3.
$$
From the fact that the spinor module is selfadjoint we infer:
$$
 \gamma^{0*}=\gamma^0,\ \gamma_5^*=\gamma_5^{\phantom{*}},\ 
 \gamma^{k*}=-\gamma^k,\ k=1,2,3.
$$
There are various matrix representations of the $\gamma$'s used in physics.
All of them respect these relations. Finally, $S$ is a supermodule. With
respect to the grading $S^+\oplus S^-$, the $\gamma$'s may be represented 
in block form: 
$$
\gamma^0=\pmatrix{0&\one\cr \one&0\cr}\,,\quad
\gamma^k=\pmatrix{0&-\sigma_k\cr\sigma_k&0\cr}\quad(k=1,2,3)\,.
$$
As these matrices indicate, the $\gamma$'s are odd operators, i.e., they map
$S^\pm$ into $S^\mp$, while $\gamma_5$ is diaonal:
$$
        \gamma_5=\pmatrix{\one & 0\cr \ 0&-\one\,\cr}\,.
$$
At the same time, the subspaces $S^\pm$ reduce
the representation of the Spin group $SL(2,\CC)$. The two subrepresentations
(so-called fundamental representations) are irreducible and inequivalent. 
Note, however, that the group $SL(2,\CC)\cong\Spin M^*_4$ is the 
{\em real spin group}, and its representation on $S$ lacks unitarity.
It would be more appropriate to pass to the complex spin group,  
$$
     \Spinc M^*_4=\exp\Big(C^2(M^*_4)\otimes\CC\Big)
                \cong SL(2,\CC)\times SL(2,\CC),
$$
whose adjoint action on $M^*_4\otimes\CC$ is known as the {\em complex Lorentz
group\/} and whose (complex) Lie algebra is
$$    
   C^2(M^*_4)\otimes\CC\cong(su(2)\oplus su(2))\otimes\CC\,,
$$ 
and regard the spinor module $S$ as representation space of this larger 
group. As both $(ie_1)(ie_2)$ and $(ie_3)e_0$ are elements in 
$$
           \Spin E_4<\Spinc M^*_4\,,
$$
so is their product, the chirality operator
$\Gamma$. Sure enough, $\Gamma$ is {\em not\/} an element of $\Spin M^*_4$
which means that a helicity change $S^\pm\to S^\mp$ cannot be effected by 
some element in $SL(2,\CC)$, but can be effected by some element in 
$$
    SU(2)\times SU(2)<SL(2,\CC)\times SL(2,\CC)\,,
$$
that is, by extension into the complex domain.

It is no surprise that the complex Lorentz group has played a 
prominent role in the Wightman formulation of field theory, especially
in the proof of the PCT theorem [11]. Moreover,
passage to the complex Lie algebra has been important for the study of 
irreducible representations of the group $SL(2,\CC)$.

\section{The Relation between Spinors and Vectors}

As part of a general folklore, spinors are thought of as ``square roots
of vectors''. In his book [4] Chevalley gave this idea a precise meaning.
Namely, in a complex setting one may argue that the tensor product
$S\otimes S^*$ recovers the vector space $\bw V\otimes\CC$. In other words,
complex $k$-vectors, i.e., elements of $\bw^kV\otimes\CC$, are sums of products
of spinors and their duals. 

The relation between spinors and vectors follows immediately from our 
route of constructing the spinor module: use complexification as the first
step and polarization as the second. In fact, the isomorphism, we want to draw
attention to, is already there and part of a large commutative diagram:
$$
\def\normalbaslines{\baselineskip30pt\lineskip5pt\lineskiplimit5pt}
\matrix{%
C(V)\otimes\CC&\mapr{\sigma}&\bw V\otimes\CC&\mapr{\ }&\bw(V_+\oplus V^*_+)\cr
      \mapd{c}&             &\mapd{\ }      &         &\mapd{\ }           \cr
      \End S  &\mapr{\ }    &S\otimes S^*   &=\!=     &
\bw V_+\otimes\bw V_+^*\cr}
$$ 
All objects of this diagram are regarded as $\ZZ_2$-graded vector spaces while
the arrows indicate isomorphisms. Typical arrows of this diagram have
previously been considered: the symbol map $\sigma$ and the Clifford action 
$c$. All other arrows relate to standard constructions in multilinear algebra 
[5].

The point of view taken by practioners in linear algebra is that $k$-vectors 
(elements of $\bw^kV$) are more fundamental than spinors (elements of $S$).
Likewise in physics, $k$-vectors are considered to be more `classical'
than spinors. One may, however, advocate the opposite point of view, namely,
that everything (vectors, tensors, operators etc.) should be built from
spinors. Such a radical chance of viewpoint has led Penrose [8] to introduce
{\em twistors\/} to reconstruct spacetime in General Relativity. This
example suggests to think of spinors in more geometric terms. It also marks the
birth of Spin Geometry. Another seminal work, that started a new line
of research, is Witten's proof [9] of the Positive-Energy Conjecture in
General Relativity where he uses spinor fields in a classical context.
Another subject, termed non-commutative geometry, has been introduced by
Connes [10] with possible applications in particle physics, not to mention
the impressive use of Dirac operators on general spin manifolds, initiated by 
Atiyah and Singer. Thanks to these developments we came to acknowledge the 
fact that the use of spinors is not only within the domain of quantum physics.

Last not least we want to point out that in Dirac theory the intimate
relationship between $k$-vectors and elements of $\End S$ has always been
re\-cog\-nized and made use of. For, if $A$ is some operator on the spinor module
$S$, it may be decomposed as $A=\sum A_k$ where $A_k$ ($k=0,\cdots,n$) are 
operators of degree $k$ and parity $(-1)^k$, obtained from complex 
antisymmetric tensors $a_{\mu_1\cdots\mu_k}$:
$$
        A_k=\frac{1}{k!}a_{\mu_1\cdots\mu_k}\sigma^{\mu_1\cdots\mu_k}
$$ where $$
    \sigma^{\mu_1\cdots\mu_k}=c\Big(\sigma^{-1}(e^{\mu_1}\wedge\cdots
    \wedge e^{\mu_k})\Big)\,.
$$
In the Dirac theory, we thus obtain
$$
  \sigma^{\mu_1\cdots\mu_k}=\frac{i^{-k}}{k!}\sum_\pi\mbox{sign}(\pi)\,
  \gamma^{\mu_{\pi(1)}}\cdots\gamma^{\mu_{\pi(k)}}\,.
$$
In short, there is a 1:1 correspondence between $a\in\bw V\otimes\CC$
and $A\in\End S$ given by $A=c(\sigma^{-1}(a))$. The parity of $a$ is
compatible with the parity of $A$, i.e., if $a\in\bw^kV$ and $(-1)^k=\pm1$, 
then $A\in\En^{\pm}V$.

\section{Universality of the Spinor Modul}

Having constructed the spinor module $S$, we immediately see that it is a
universal object within the category of complex irreducible Clifford modules.
That is to say, any irreducible Clifford action $C(V)\otimes\CC\to\End E$
factorizes,
$$
      C(V)\otimes\CC\arr{c}\End S\arr{r}\End E\,,
$$ 
the algebraic isomorphism $r$ being induced by some vector space isomorphism:
$$         
        S\arr{g}E,\quad r(b)=gbg^{-1},\qquad b\in\End S\,.
$$
Therefore, the spinor module is, up to isomorphism, the only irreducible
complex Clifford module. 

The problem of determining all possible Clifford modules $E$ reduces to
the study of representations 
\begin{equation}
    \label{repr}
                   r:\End S\to\End E\,.        
\end{equation}
Unless $\dim S=1$, one cannot construct a trivial representation (some 
homomorphism $\End S\to\CC$) of the algebra $\End S$. Since 
$\dim S=2^{n/2}\ge2$, trivial representations of the Clifford algebra do not 
occur. More is true. Simple facts about endomorphismen algebras teach us
that any representation (\ref{repr}) is {\em completely reducible\/} and 
decomposes into subrepresentations, each one of them being isomorphic to the 
fundamental representation ($\id:\End S\to\End S$). A representation of an 
algebra (or group) is called {\em primary}, if it is a multiple of a single 
irreducible representation. Summarizing, {\em any Clifford module is 
primary\/} or, equivalently, assumes the form
$$
              E\cong W\otimes S
$$
where it is understood that $c(a)$ acts trivially on the vector space $W$:
$$
   c(a)(w\otimes s)=w\otimes as\qquad (w\in W,\,s\in S,a\in C(V)\otimes\CC).
$$ 
Modules with this structure are called {\em twisted spinor modules}.
The space $W$ is termed {\em twisting space}. Abstractly, the space $W$ can
be identified with $\Hom_{\rm Cl}\,(S,E)$, i.e., with the vector space
of linear maps $w:S\to E$ that commute with the Clifford action. 
The isomorphism
$$
     \Hom_{\rm Cl}\,(S,E)\otimes S\to E,\qquad w\otimes s\mapsto ws
$$
is then obvious. 

Conversely, let $W$ be any vector space. Then the tensor product
$W\otimes S$ is a Clifford module (with trivial Clifford action on $W$).
In physics this construction is used to incorporate further degrees of
freedom beyond those of the spin polarization. These extra degrees may
describe the momentum of a particle (see Dirac's theory of the electron
for example) or be related to internal symmetries as is the case in
gauge theories. The above result ``Clifford modules are twisted spin
modules'' imposes severe restrictions on model building.

Whatever the twisting space $W$, the module $E=W\otimes S$ is $\ZZ_2$-graded, 
hence a supermodule:
$$
          E=E^+\oplus E^-,\qquad E^\pm=W\otimes S^\pm\,.
$$
This gives $\End E$ the structure of a superalgebra. We will, however,
also deal with cases where the twisting space carries a $\ZZ_2$-grading.
Then
$$
         E^\pm=(W^+\otimes S^\pm)\oplus(W^-\otimes S^\mp)
$$
and even(odd) operators on $S$ extend to even(odd) operators on $E$.
This shows that in a variety of situations the twisted spinor module is
indeed a supermodule.

Recall that we have
previously constructed a canonical scalar product in the spinor module $S$.
The existence of another scalar product in the twisting space $W$ would turn
the module $E$ into a Hermitian space with a selfadjoint Clifford action,
a situation we normally encounter in physical applications.

\section{Supercommuting Endomorphisms}

This section is devoted to studying the structure of $\End E$ when $E$
is some complex $C(V)\otimes\CC$ supermodule and $V$ is pseudo-Euclidean.
To facilitate the discussion we use a shorthand for the complexified
Clifford algebra: 
$$        A=C(V)\otimes\CC\,.$$ 
The main result will be the isomorphism
\begin{equation}
      \End E\cong A\skt\En_A\,E\,.      \label{EndE}
\end{equation}
Since both $A$ and $\En_A\,E$ are superalgebras, we have to exercise
some care:
\begin{enumerate}
\item The algebra $\En_A\,E$ has endomorphisms $b$ as its elements that
{\em supercommute\/} with the Clifford action: $\sco{a,b}=0$, $a\in A$.
\item The tensor product $\skt$ is special for $\ZZ_2$-graded algebras (often
called the {\em skew tensor product\/}): 
$$
   (a\skt b)(a'\skt b')=\cases{-aa'\skt bb' & if $a,b$ are odd\cr
                       \phantom{-}aa'\skt bb' & otherwise.\cr}
$$
Sure enough, as tensor product of vector spaces there would be no difference 
and hence no confusion.
\end{enumerate}
The proof of (\ref{EndE}) runs as follows. 
Choose some orthogonal basis $(e_i)_{i=1}^n$ in
$V\otimes\CC$, and consider the induced basis $e_I$ in $A$.
For $k=0,1,\ldots,n$ and $I\subset\{1,\ldots,k\}$ we define projection 
operators $P_I^{(k)}\in\End A$ recursively. If $k=0$ we put 
$P_\emptyset^{(0)}=\id$ and, for $k\ge1$,
\begin{equation}
  \label{PI}
  P_I^{(k)}a =\cases{ -\hf\sco{e_k,e_kP^{(k-1)}_Ia} & if $k\nein I$\cr
   -\hf e_k\sco{e_k,P^{(k-1)}_{I\backslash\{k\}}a} & if $k\in I$  \cr}
  \qquad(a\in A).
\end{equation}
Using the quantization map (of Section 4) one sees that each supercommutator 
$-\hf\sco{e_k,\cdot}$ acts like an annihilation (contraction) operator 
while each multiplication from the right by $e_k$ acts like a creation 
operator. So one obviously deals here with the product of these two
operators, either in normal or in reversed order. In the physics literature,
such products occur as `number operators' having eigenvalues 0 and 1 only,
so they may justly be called projection operators, too.
The correspondence with Fermi operators helps to establish the following 
results.
For $J\subset\{1,\ldots,n\}$,
$$
   P_I^{(k)}e_J =n_k(I,J)e_J,\qquad n_k(I,J)
   =\cases{1 & if $\{1,\ldots,k\}\subset (I\cap J)\cup(I^c\cap J^c)$\cr
           0 & otherwise.\cr}
$$
($I^c=\,$compliment of $I$ in $\{1,\ldots,n\}$). For $P_I=P^{(n)}_I$ we have 
\begin{equation}
    \label{PHut}
   P_I= e_IQ_I, \qquad Q_Ie_J=\delta_{IJ}1,\qquad \sum_I P_I=\id    
\end{equation}
with operators $Q_I\in\End A$, serving as left inverses of the
multiplication by $e_I$. Explicitly, for a suitable choice of $\sigma_I=\pm1$,
$$
   Q_Ia =(-\hf)^n\sigma_I
        \sco{e_n,\ldots\sco{e_2,\sco{e_1,e_{I^c}\,a}}\ldots}\qquad (a\in A).
$$
Owing to the existence of these operators, the Clifford algebra is seen to be 
{\em simple}. Namely, by virtue of the relations (\ref{PHut}), any nontrivial 
ideal must contain the unit $1\in A$ and hence coincide with the whole of $A$.
As a consequence, Clifford actions are always injective.

Consider now a Clifford module $E$. Under the algebraic homomorphism
$c:A\to\End E$, the operators $P_I^{(k)}$ and $Q_I$ have images 
$\hat{P}_I^{(k)}$ and $\hat{Q}_I$ acting linearly on $\End E$. 
By analogy to (\ref{PI}), setting $c_k=c(e_k)$ and 
$\hat{P}_\emptyset^{(0)}=\id$, we recursively define
\begin{equation}
  \label{hPI}
  \hat{P}_I^{(k)}b 
  =\cases{ -\hf\sco{c_k,c_k\hat{P}^{(k-1)}_Ib} & if $k\nein I$\cr
   -\hf c_k\sco{c_k,\hat{P}^{(k-1)}_{I\backslash\{k\}}b} & if $k\in I$  \cr}
  \qquad(b\in\End E).
\end{equation}
Similarly, setting $c_I=c(e_I)$, we also define
$$
   \hat{Q}_Ib =(-\hf)^n\sigma_I
        \sco{c_n,\ldots\sco{c_2,\sco{c_1,c_{I^c}\,b}}\ldots}
$$ 
so that the operator identities $\hat{P}_I=c_I\hat{Q}_I$ hold.   
We show now the validity of the completeness relation 
$$
         \sum_I\hat{P}_Ib=b\qquad (b\in \End E).
$$ 
Though, by construction, this relation is certainly valid for $b=c(a)$ 
and $a\in A$ arbitrary, we cannot assert that the relation extends 
{\em automatically\/} to all of $\End E$. But in fact it does.
To make the proof more transparent, let us introduce auxiliary elements
$$
                b_k=\sum \hat{P}^{(k)}_Ib\in \End E\qquad(k=0,1,\ldots,n)
$$
where the sum is over $I\subset\{1,\ldots,k\}$. The assertion may now be
stated as $b_n=b$, and what is known may be stated as 
$b_0=\hat{P}^{(0)}_\emptyset b =b$. Therefore, the proof would be complete,
once we have shown that $b_k$ is independent of $k$.

The key equation is
\begin{equation}
  \label{hfs}
     -\hf\sco{c_k,c_kb}-\hf c_k\sco{c_k,b}=b\qquad (b\in\End E)  
\end{equation}
which follows from the generalized Jacobi identity and
$$
       \sco{c_k,c_k}b=c(\sco{e_k,e_k})\,b=-2c(1)b=-2b\,.
$$
From (\ref{hPI}) and (\ref{hfs}) we infer
$$
      \hat{P}^{(k)}_Ib+\hat{P}^{(k)}_{I\cup\{k\}}b=\hat{P}^{(k-1)}_Ib\,,
      \qquad I\subset\{1,\ldots,k-1\}.
$$
Summing both sides over $I$ yields $b_k=b_{k-1}$ which completes the proof.

Any $b\in\End E$ can now be decomposed as 
$$
    b=\sum_I\hat{P}_Ib=\sum_Ic(e_I)\hat{Q}_Ib=\sum_Ie_I\skt\hat{Q}_Ib\,.
$$
This way we have explicitly constructed the decomposition in $A\skt\En_A\,E$. 
It remains to demonstrate the property
\begin{equation}
      \sco{c(a),\hat{Q}_Ib}=0\qquad (a\in A)\label{skew}
\end{equation}
stating that $\hat{Q}_Ib$ is some element of $\En_A\,E$ and 
also that the tensor product is `skew'. Reason: skewness implies that, for 
arbitrary $b\in\En_A\,E$,
$$
  \sco{c(a),b}=(a\skt 1)(1\skt b)\pm(1\skt b)(a\skt 1) = 0
$$
with a plus sign if both $a$ and $b$ are odd and a minus sign otherwise.
It suffices to prove the property (\ref{skew}) for the special case $a=e_i$.
To see the strategy of the proof, take $a=e_n$ first:
\begin{eqnarray*}
\sco{c_n,\hat{Q}_Ib}&\sim&
\sco{c_n,\sco{c_n,\ldots,\sco{c_1,c_{I^c}}b}\ldots}\\
&=&\sco{\sco{c_n,c_n},\sco{c_{n-1},
\ldots}}-\sco{c_n,\sco{c_n,\sco{c_{n-1},\ldots}}}\\
&=&-\sco{c_n,\sco{c_n,\ldots,\sco{c_1,c_{I^c}\,b}}}\ =\ 0\,.
\end{eqnarray*}
To pass from the first to the second line we have used the generalized Jacobi 
identity. To obtain the third line we have used $-\hf\sco{c_n,c_n}=\id$.
Suppose now that we run the same calculation with $a=e_i$. We would
also apply the Jacobi identity and the property 
$-\hf\sco{c(w),c(w')}=(w,w')\id$, valid for all $w,w'\in V\otimes\CC$. 
In effect, we make use of the formula
$$
    \sco{c(w),\sco{c(w'),\cdot}}+\sco{c(w),\sco{c(w'),\cdot}}=0
$$
to carry out the proof along the same lines. So far, Eq.(\ref{skew}) is seen
to be correct for the generators $a=e_i$ only. But the general 
assertion (for $a=e_I$) is an immediate consequence.

Last not least we draw attention to the fact that all algebras involved in
the foregoing discussion are $\ZZ_2$-graded. It is easily checked that
the decomposition (\ref{EndE}) conforms to these structures, i.e., we have
$$
\En^\pm\,E =(A^+\skt\En_A^\pm\, E)\oplus(A^-\skt\En_A^\mp\, E)\ .
$$
where $A=C(V)\otimes\CC$ as before.
\vspace{15mm}\par\noindent
{\Large\bf References}\\[2mm]
\begin{enumerate}
\item Vehns, Ch. Diploma Thesis, RWTH Aachen 1999
\item Roepstorff, G.: Superconnections: An Interpretation of the Standard 
      Model, hep-th/9907221, to appear in El.J.Diff.Eq. (USA).
\item Berline, N., E.\ Getzler, and M.\ Vergne: {\em Heat Kernels and Dirac
      Operators}, Springer, Berlin Heidelberg 1992
\item Chevalley, C.: {\em The algebraic theory of spinors}, Columbia 
      University Press, New York 1954
\item Greub, W.: {\em Multilinear Algebra}, 2nd Ed.\ Springer, 
      New York 1978
\item Lawson, H.B.\ and M.-L.\ Michelson: {\em Spin Geometry}, Princeton
      Mathematical Series {\bf 38}, Princeton Univ.\ Press, Princeton 1989
\item Thaller, B.: {\em The Dirac Equation}, Springer, Berlin Heidelberg 1992
\item Penrose, R.\ and R.\ Rindler: {\em Spinors and Space-Time}, Cambridge
      Univ.\ Press, Cambridge UK 1984
\item Witten E.: A new Proof of the Positive Energy Theorem,
      Commun.\ Math.\ Phys.\ {\bf 80}, 381-402 (1981)
\item Connes, A.: {\em Noncommutative differential geometry and the structure
      of space-time}, Proceedings of the Symposium on Geometry, Huggett, S.A.\
      (ed.) et al., pp.49-80, Oxford Univ.\ Press, Oxford UK 1998
\item Streater, R.F.\ and A.S.\ Wightman: {\em PCT, Spin \& Statistics, and
      all That}, Mathematical Physics Monograph Series, Benjamin, New York 1964
\end{enumerate}
\end{document}